\documentclass[superscriptaddress,pre,showpacs,preprintnumbers,amsmath,amssymb,aps]{revtex4}
\usepackage{graphics}
\usepackage{graphicx}
\usepackage{epsfig}
\usepackage{amssymb}
\usepackage{color}
\usepackage{dcolumn}%
\usepackage{bm}% bold math

\begin{document}

\title{Enhancing heat transfer in nanofluids by carbon nanofins}% Force breaks \\

\author{Eliodoro Chiavazzo}\email{eliodoro.chiavazzo@polito.it}
%\affiliation{Department of Energetics, Politecnico di Torino,
%Corso Duca degli Abruzzi 24, 10129 Torino, Italy}

\author{Pietro Asinari}\email{pietro.asinari@polito.it}

\affiliation{Department of Energetics, Politecnico di Torino,
Corso Duca degli Abruzzi 24, 10129 Torino, Italy}

\date{\today}% It is always \today, today,

\begin{abstract}
%In this work, we focus on the usage of carbon nanotubes as nanofins to enhance heat transfer between a surface and a fluid in contact with it. To this end, we first investigate the thermal conductivity of the latter nanostructures by means of classical non-equilibrium molecular dynamics simulations using the package GROMACS. Next, thermal conductance at the interface between a single wall carbon nanotube (nanofin) and water molecules is computed by means of both steady-state and transient molecular dynamics.
Nanofluids are suspensions of nanoparticles and fibers which have recently attracted much attention due to their superior thermal properties. Here, nanofluids are studied in the sense of nanofins transversally attached to a surface, so that dispersion within a fluid is mainly dictated by design and manufacturing processes.
%Text for this section of the abstract \ldots
We focus on single carbon nanotubes thought as nanofins to enhance heat transfer between a surface and a fluid in contact with it. To this end, we first investigate the thermal conductivity of those nanostructures by means of classical non-equilibrium molecular dynamics simulations. Next, thermal conductance at the interface between a single wall carbon nanotube (nanofin) and water molecules is assessed by means of both steady-state and transient numerical experiments.			
%Text for this section of the abstract \ldots
Numerical evidences suggest a pretty favorable thermal boundary conductance (order of $10^{-7}$ $[W m^{-2} K^{-1}]$) which makes carbon nanotubes ideal candidates for constructing nanofinned surfaces.
\end{abstract}

%\begin{keyword}
%Multi-scale dynamical system \sep Combustion \sep Invariant manifold
%\end{keyword}

\pacs{66.70.+f,~05.10.-a,~62.25.+g,~68.35.Md}

\maketitle

\section{Background}
Nanofluids are suspensions of nanometer-sized solid particles and fibers, which have recently become a subject of growing scientific interest because of reports of greatly enhanced thermal properties \cite{Wang2010,Lee2007}. Filler dispersed in a nanofluid is typically of nanometer size, and it has been shown that such nanoparticles dispersed in a base fluid are able to endow it with a much higher effective thermal conductivity than pure fluid \cite{Hwang2005, Assael2006}: Significantly higher than those of commercial coolants such as water and ethylene glycol. In addition, nanofluids show an enhanced thermal conductivity compared to theoretical predictions based on the Maxwell equation for a well-dispersed particulate composite model. These features are highly desirable for applications, and nanofluids may be a strong candidate for new generation of coolants \cite{Lee2007}. The use of nanofluids in many industrial sectors, including energy supply and production, transportation and electronics appears promising. A review about experimental and theoretical results on the mechanism of heat transfer in nanofluids can be found in Ref. \cite{Terekhov2010}, where Authors discuss issues related to the technology of nanofluid production, experimental equipment, and features of measurement methods. A large degree of randomness and scatter has been observed in the experimental data published in the open literature. Given the inconsistency in these data, it is impossible to develop a comprehensive physical-based model that can predict all the trends. This also points out the need for a systematic approach in both experimental and theoretical studies \cite{Bahrami2007}.

In particular, carbon nanotubes (CNTs) have attracted great interest for nanofluid applications, because of the claims about their exceptionally high thermal conductivity \cite{Berber00PRL}. However, recent experimental findings about CNTs report an anomalously wide range of enhancement values that continue to perplex the research community and remain unexplained \cite{Venkata2008}. For example, some experimental studies showed that there is a modest improvement in thermal conductivity of water at a high loading of multi-walled carbon nanotubes (MW-CNTs), $\sim$35\% increase for a 1 wt\% MWNT nanofluid \cite{Acchione2006}. These authors attribute the increase to the formation of a nanotube network with a higher thermal conductivity. On the contrary, at low nanotube content, $<$0.03 wt\%, they observed a decrease in thermal conductivity upon an increase of nanotube concentration. On the other hand, more recent experimental investigations showed that the enhancement of thermal conductivity as compared to water is varying linearly when MW-CNT weight content is increasing from 0.01 to 3 wt\%. For a MWNT weight content of 3 wt\% the enhancement of thermal conductivity reaches 64\% of the base fluid (e.g. water). The average length of the nanotubes appears to be a very sensitive parameter. The enhancement of thermal conductivity compared to water alone is enhanced when nanotube average length is increasing in the 0.5-5 $\mu$m range \cite{Glory2008}.

Clearly, there are difficulties in the experimental measurements \cite{Choi2009}, but the previous results also reveal some underlaying technological problems. First of all, the CNTs show some bundling or the formation of aggregates originating from the fabrication step. Moreover it seems reasonable that CNTs encounter poor dispersibility and suspension durability due to the aggregation and surface hydrophobicity of CNTs as a nanofluid filler. Therefore, the surface modification of CNTs or additional chemicals (surfactants) have been required for stable suspensions of CNTs, because the base fluid for the coolant has polar characteristics. In the case of surface modification of CNTs, water-dispersible CNTs have been extensively investigated for potential applications, such as biological uses, nanodevices, novel precursors for chemical reagents, and nanofluids \cite{Lee2007}. A popular solution for increasing dispersion of CNTs is based on functionalization. Oxygen-containing functional groups have been introduced on the CNT surfaces and more hydrophilic surfaces have been formed during this treatment, which enabled to make stable and homogeneous CNT nanofluids \cite{Xie2003}. Alternative solutions relay on ultrasonic disrupting, which significantly decreases the size of agglomerated particles and number of primary particles in a particle cluster, such that thermal conductivity increases with the elapsed ultrasonication time \cite{Amrollahi2008}. In any case, it is clear that many parameters affect the thermal conductivity including size, shape and source of nanotubes, surfactants, power of ultrasonic, time of ultrasonication, elapsed time after ultrasonication, pH, temperature, particle concentration and surfactant concentration \cite{Meibodi2010}. Hence, there is a lot of room for technological optimization.

In the above brief review, only the free suspensions of CNTs have been considered, i.e. nanofluids were the highly thermally conductive filler is free to move. In fact, beyond the favorable aspect ratio, the CNTs dispersed in nanofluids lead to enhanced thermal conductivity, which cannot be explained by traditional conductivity theories such as the Maxwell's mixing theory or other macroscale approaches. Recently, Jang \& Choi \cite{JangChoi2004} have found that the Brownian motion of nanoparticles at the molecular and nanoscale level is a key mechanism governing the thermal behavior of nanoparticle-fluid suspensions. They have devised a theoretical model that accounts for the fundamental role of dynamic nanoparticles in nanofluids. Essentially, these authors discovered a fundamental difference between solid/solid composites and solid/liquid suspensions in size-dependent conductivity. In the original reference \cite{JangChoi2004}, they claim that, even though the random motion of nanoparticles is characterized by zero time average, the vigorous and relentless interactions between liquid molecules and nanoparticles at the molecular and nanoscale level translate into conduction at the macroscopic level, because there is no bulk flow.

However there is another way to exploit the concept of nanofluids, i.e. to modify the base fluid properties by nanostructures. Essentially one promising way consists in fixing the relative position of the CNTs on nanoengineered surfaces. This means moving from \textit{nanotubes} (where the focus is on the shape of the nanostructure) to \textit{nanofins} (where the focus is on the function of the nanostructure). In this way, the Brownian motion of nanoparticles is gone, but there is still the possibility to enhance the heat transfer by increasing the effective cooling surface area. The fundamental technological advantage is that the dispersion of CNTs is controlled by design and only limited by the manufacturing process. Nowadays efficient cooling of silicon chips using microfin structures made of aligned MW-CNT arrays has been achieved \cite{Kordas2007}. The tiny cooling elements mounted on the back side of the chips enable power dissipation from the heated chips on the level of modern electronic devices demands. Nanotubes utilized as thermal fins (nanofins) are mechanically superior compared to other materials being ten times lighter, flexible, and stiff at the same time \cite{Kordas2007}. Nanofins are extensively investigated also from a modeling standpoint \cite{Singh2009}. The current challenge is to develop industrial manufacturing processes for macroscopic growth of carbon nanotube mats \cite{Musso2007}.

The present paper aims to investigate by molecular mechanics based on force fields (MMFF) the thermal performance of nanofins made of single wall CNTs (SW-CNTs) cooled by water. This work focuses on the astonishing thermal properties of these nanostructures, in particular, when they interact with the surrounding base fluid. The single wall CNTs were selected mainly on the basis of time constraints due to our parallel computational facilities. The following analysis can be split into two parts. First of all, the heat conductivity of SW-CNTs is estimated numerically by both simplified model (Section 1, where this approach is proved to be inadequate) and detailed three-dimensional model (Section 2). This first step is required for tuning the numerical model and validate the vacuum results with literature results. Next, the thermal boundary conductance between SW-CNT and water is computed by two methods: the steady state method (Section 3.1), mimicking ideal cooling due to strong Brownian motion, and transient method (Section 3.2), taking into account only weak Brownian motion.
%\red{Text for this section \ldots}

\section{Heat conductivity of single-wall carbon nanotubes: A simplified model}
In order to significantly downgrade the difficulty of studying energy transport processes within a carbon nanotube, authors often resort to simplified low dimensional systems such as one-dimensional lattices \cite{Savin03,Kaburaki93,LiuLi07,LINIAN,Musser10,Wang07PRL}. In particular, heat transfer in a lattice is typically modeled by the vibrations of lattice particles interacting with the nearest neighbors and by a coupling with thermostats at different temperatures. The latter are the popular {\em numerical experiments} based on non-equilibrium molecular dynamics (NEMD). In this respect, to the end of measuring the thermal conductivity of ($5,5$) single wall nanotubes (SWNT), we set up a model for solving the equations of motion of the particle chain pictorially reported in Fig. 1 where each particle represents a ring with 10 atoms in the real nanotube. In the present model, Carbon-Carbon bonded interactions between {\em first neigbors} (i.e. atoms of the $i$th particles and atoms of the particles $i \pm 1$) separated by a distance $r$ are taken into account by a Morse-type potential expressed in terms of deviations $x=r-r_0$ from the bond length $r_0$:
\begin{equation}\label{Morse1D}
V_{b} \left( x \right) = V_0 \left( e^{-2\frac{x}{a}} -2 e^{-\frac{x}{a}} \right),
\end{equation}   
where $V_0$ is the bond energy while $a$ is assumed $a=r_0/2$. Following \cite{Brenner02}, bond energy is $V_0=4.93[eV]$, while the distance between two consecutive particles at equilibrium is assumed $r_0=0.123 [nm]$. At any arbitrary configuration the total force, $F_i$, acting on the $i$th particle is computed as:
\begin{equation}\label{inner.force}
F_i = - N_{bon} \sin{\vartheta} \left[ \frac{\partial V_b}{\partial x}\left(dx_{i-1}\right) + \frac{\partial V_b}{\partial x}\left(dx_{i+1}\right) \right],
\end{equation}
with $dx_{i - j}=x_i-x_{i - j}$, $dx_{i + j}=x_{i + j} - x_i$ and $N_{bon}$ denoting the number of Carbon-Carbon bonds between two particles, whereas a penalization factor $\sin{\vartheta}$ can be included to account for bonds not aligned with the tube axis.
Here, we use free-end boundary condition, hence forces experienced by particles at the ends of the chain read:
\begin{equation}
F_1 = - N_{bon} \sin{\vartheta} \left[ \frac{\partial V_b}{\partial x}\left(dx_{2}\right) \right], \quad F_N = - N_{bon} \sin{\vartheta} \left[ \frac{\partial V_b}{\partial x}\left(dx_{N-1}\right) \right].
\end{equation}
Let $p_i$ and $m_i$ be the momentum and mass of the $i$th particle, respectively, the equations of motion for inner particles take the form:
\begin{equation}\label{free.motion.1D}
\frac{dx_i}{dt} = \frac{p_i}{m_i}, \quad \frac{dp_i}{dt} = F_i,
\end{equation}
whereas the outermost particles ($i=1,N$) are coupled to Nos\'e-Hoover thermostats and are governed by the equations:
\begin{equation}\label{nh.motion.1D}
\frac{dx_i}{dt} = \frac{p_i}{m_i}, \quad \frac{dp_i}{dt} = F_i - \xi p_i, \quad \frac{d\xi}{dt} = \frac{1}{Q} \left[ {\frac{p_i^2}{2m_i}} - N_f k_b T_0\right], \quad Q=\frac{\tau_T^2 T_i}{4 \pi^2},
\end{equation} 
with $k_b$, $T_0$, $N_f$ and $\tau_T$ denoting the Boltzmann constant, the thermostat temperature, number of degrees of freedom and relaxation time, respectively, while the auxiliary variable $\xi$ is typically referred to as {\em friction coefficient} \cite{hoover03}. Nos\'e-Hoover thermostatting is preferred since it is deterministic and it preserves canonical ensemble (see, e.g., \cite{Hunen05} and \cite{Frenkelbook} for further details on thermostats in molecular dynamics simulations).

Local temperature $T_i(t)$ at a time instant $t$ is computed for each particle $i$ using energy equipartition:
\begin{equation}
T_i(t)=\frac{1}{k_b N_f} \left\langle \frac{p_i(t)^2}{m_i} \right\rangle,
\end{equation}
where $\left\langle{ }\right\rangle$ denotes time averaging. On the other hand, local heat flux $J_i$ transferred between particle $i$ and $i+1$, can be linked to mechanical quantities by the following relationship \cite{LiuLi07,Musser10}:
\begin{equation}\label{mech.heat.flux}
J_i=\left\langle { \frac{p_i}{m_i} \frac{\partial V_b}{\partial x}\left(dx_{i+1}\right) } \right\rangle.
\end{equation}

The above simplified model has been tested in a range of {\em low} temperature ($300[K]<T<1000[K]$), where we noticed that it is not suitable to predict normal heat conduction (Fourier's law). In other words, at steady state (i.e. when heat flux is uniform along the chain and constant in time), it is observed a finite heat flux although no meaningful temperature gradient could be established along the chain (see Fig. 1). Thus, the above results predict a divergent heat conductivity. Here, it is worth stressing that one-dimensional lattices with harmonic potentials are known to violate Fourier's law and exhibit a flat temperature profile and divergent heat conductivity. On the other hand, consistently with the present numerical experiments, it has been demonstrated that anharmonicity alone is insufficient to ensure normal heat conduction \cite{Savin03}.   
%%%%%%%%%%%%%%%%%%%%%%
\begin{figure}
	\centering
		\includegraphics[width=0.55\textwidth]{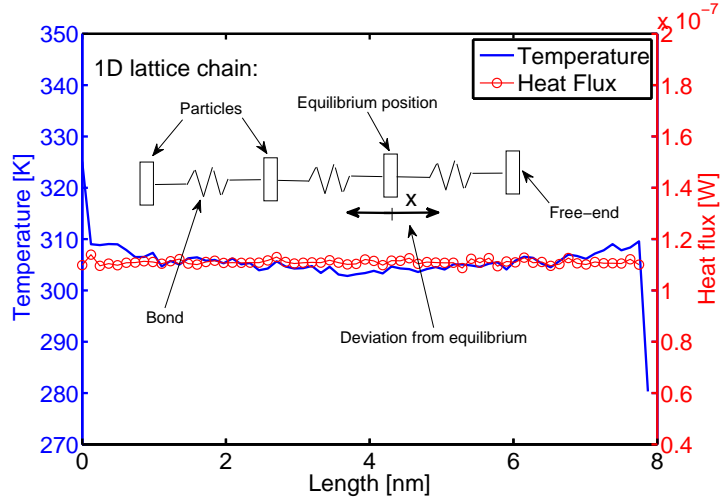}
	\caption{One-dimensional model: Lattice chain of particles interacting according to a Morse-type potential (\ref{Morse1D}). End-particles are coupled to Nos\'e-Hoover thermostats at different temperature ($T_{hot}=320[K]$ and $T_{cold}=280[K]$). Despite of the anharmonicity of the potential, normal heat conduction (Fourier's law) could not be established. Here, heat flux is computed by eq. (\ref{mech.heat.flux}). However, consistent results are obtained by eq. (\ref{heat.flux.nh}) below which predicts: $\left\langle \xi_{hot} \right\rangle k_b T_{hot} = - \left\langle \xi_{cold} \right\rangle k_b T_{cold} = 1.11 \times 10^{-7} [W]$.}\label{1Dchain}
\end{figure}
%%%%%%%%%%%%%%%%%%%%%%

\section{Heat conductivity of single-wall carbon nanotubes: Detailed three dimensional models}
In all simulations below, we have adopted the open-source molecular dynamics (MD) simulation package GROningen MAchine for Chemical Simulations (GROMACS) \cite{gromacs01,gromacs02,gromacsweb} in order to investigate the energy transport phenomena in three-dimensional SWNT obtained by a freely available structure generator (Tubegen) \cite{tubegen}. Three harmonic terms are used to describe the Carbon-Carbon bonded interactions within the SWNT. Namely, a bond stretching potential (between two covalently bonded carbon atoms $i$ and $j$ at a distance $r_{ij}$):
\begin{equation}\label{bond.potentials.st}
V_b \left( r_{ij} \right)=\frac{1}{2} k_{ij}^b \left( r_{ij} - r_{ij}^0 \right)^2,
\end{equation}
a bending angle potential (between the two pairs of covalently bonded carbon atoms ($i,j$) and ($j,k$))
\begin{equation}\label{bond.potential.an}
V_a \left( \theta_{ijk} \right) = \frac{1}{2} k_{ijk}^\theta \left( \cos \theta_{ijk}  - \cos\theta_{ijk}^0 \right)^2,
\end{equation}
and the Rychaert-Bellemans potential for proper dihedral angles (for carbon atoms $i$, $j$, $k$ and $l$)
\begin{equation}\label{bond.potential.tor}
V_{rb} \left( \phi_{ijkl} \right) = \frac{1}{2} k _{ijkl}^\phi \left( 1- \cos{2 \phi_{ijkl}} \right)
\end{equation}
are considered in the following MD simulations. Here, $\theta_{ijk}$ and $\phi_{ijkl}$ represent all the possible bending and torsion angles, respectively, while $r_{ij}^0=0.142 [nm]$ and $\theta_{ijk}^0 = 120^\circ$ are reference geometry parameters for graphene. Nonbonded Van der Waal interaction between two individual atoms $i$ and $j$ at a distance $r_{ij}$ can be also included in the model by a Lennard-Jones potential:
\begin{equation}\label{LJ.pot}
V_{nb}=4 \epsilon_{CC} \left[ \left( \frac{\sigma_{CC}}{r_{ij}} \right)^{12} - \left( \frac{\sigma_{CC}}{r_{ij}} \right)^6 \right],
\end{equation}
where the force constants $k_{ij}^b$, $k_{ijk}^\theta$ and $k_{ijkl}^\phi$ in (\ref{bond.potentials.st}), (\ref{bond.potential.an}), (\ref{bond.potential.tor}) and parameters ($\sigma_{CC}$, $\epsilon_{CC}$) in (\ref{LJ.pot}) are chosen according to the table 1 below (see also \cite{Guo91} and \cite{Koum01}).
%%%%%%%
In reversible processes, differentials of heat $dQ_{rev}$ are linked to differentials of a state function, entropy, $ds$ through temperature: $dQ_{rev}=T ds$. 
%%%%%%%
Moreover, following Hoover \cite{hoover94,hoover03}, entropy production of a Nos\'e-Hoover thermostat is proportional to the time average of the friction coefficient $\left\langle \xi \right\rangle$ trough the Boltzmann constant $k_b$ hence,
once a steady state temperature profile is established along the nanotube, the heat flux per unit area within the SWNT can be computed as: 
%%%%%%%%%%%
\begin{equation}\label{heat.flux.nh}
q = - \left\langle \xi \right\rangle \frac{N_f k_b T}{S_A},
\end{equation}
%%%%%%%%%%%
where the cross section $S_A$ is defined as $S_A = 2 \pi r b$, with $b=0.34[nm]$ denoting the Van der Waals thickness (see also \cite{Shelly10}). Here, the use of formula (\ref{heat.flux.nh}) is particularly convenient since the quantity $\left\langle \xi \right\rangle$ can be readily extracted from the output files in GROMACS.
%%%%%%%%%%%%%%%%%%%%%%
\begin{figure}
	\centering
		\includegraphics[width=0.65\textwidth]{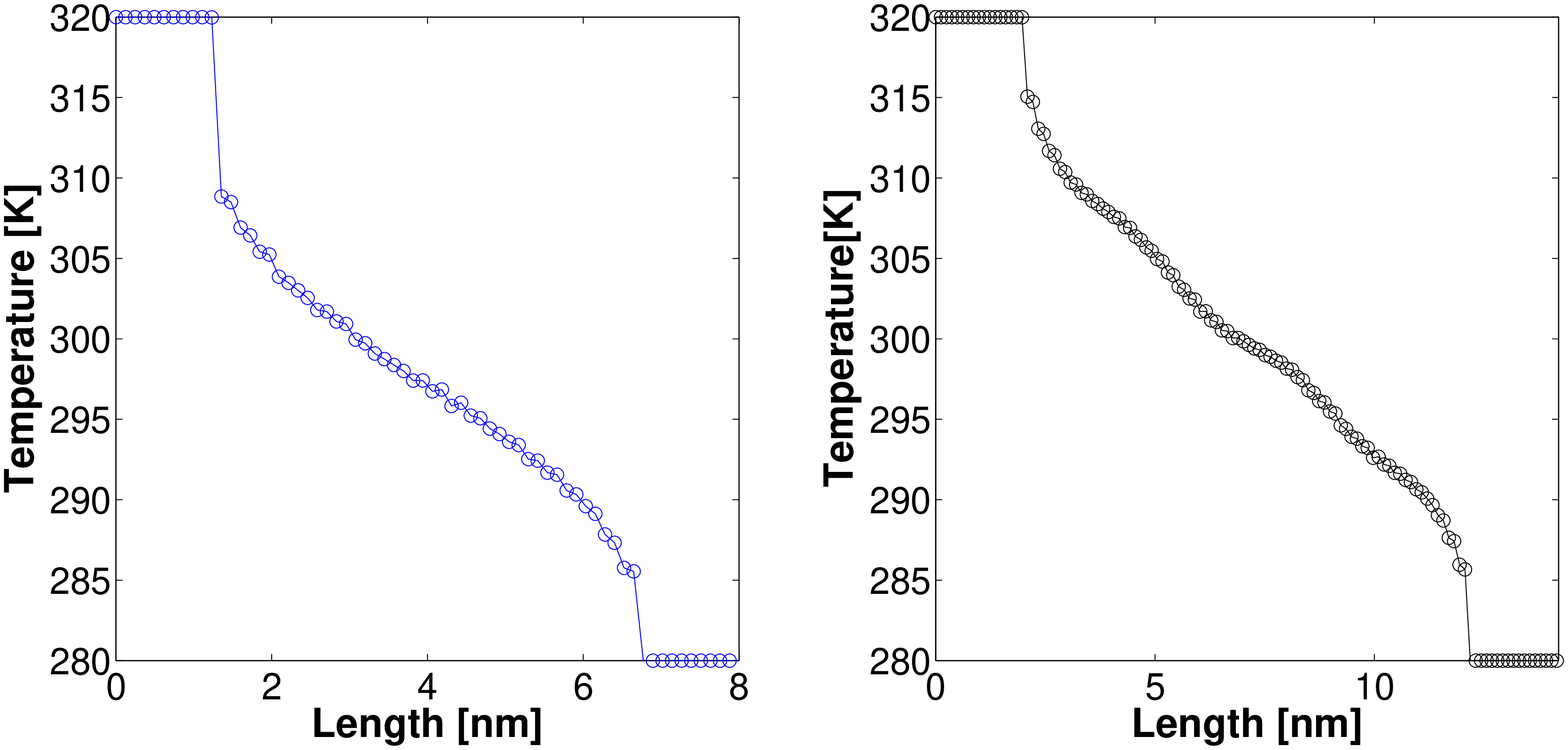}
		\includegraphics[width=0.45\textwidth]{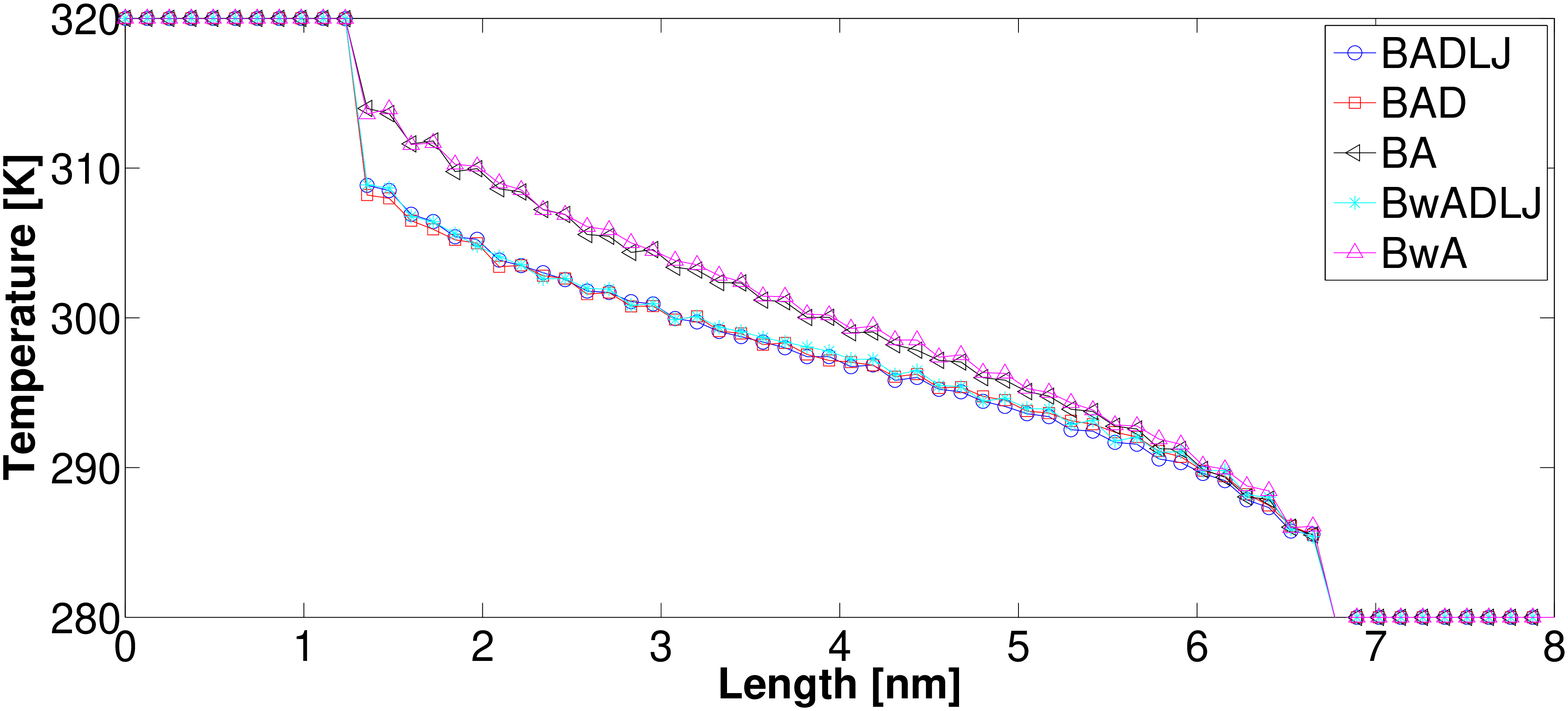}
	\caption{Three-dimensional model: Nos\'e-Hoover thermostats are coupled to the end atoms of a ($5,5$) SWNT. (Top) Both bonded (\ref{bond.potentials.st}) (\ref{bond.potential.an}) (\ref{bond.potential.tor}) and nonbonded interactions (\ref{LJ.pot}) are considered. In a three-dimensional structure, harmonic bonded potentials do give rise to normal heat conduction. Temperature profiles for two lengths ($5.5[nm]$ and $10[nm]$) are reported. (Down) Several setups have been tested where some of the interaction potentials (\ref{bond.potentials.st}), (\ref{bond.potential.an}), (\ref{bond.potential.tor}) and (\ref{LJ.pot}) are dropped out. BADLJ: $V_b$, $V_{an}$, $V_{rb}$ and $V_{nb}$ are consedered. BAD: $V_b$, $V_{an}$, $V_{rb}$ are considered. BA: $V_b$ and $V_{an}$ are considered. Bw denotes that $V_b$ is computed with a smaller force constant $k_{ij}^b = 42000 [kJ mol^{-1} nm^{-2}]$ according to \cite{Brenner02}.}\label{3DSWNT}
\end{figure}
%%%%%%%%%%%%%%%%%%%%%%

The measure of both the slope of temperature profile along the inner rings of SWNT in Fig. 2 and Fig. 3 and heat flux by (\ref{heat.flux.nh}) enables us to evaluate heat conductivity $\lambda$ according to Fourier's law. It's worth stressing that, as shown in the latter figures, unlike one-dimensional chains such as the one discussed above, fully three-dimensional models do predict normal heat conduction even when using harmonic potentials as (\ref{bond.potentials.st}), (\ref{bond.potential.an}) and (\ref{bond.potential.tor}).
%%%%%%%%%%%
Interestingly, in our simulations it is possible to drop out at will some of the interaction terms $V_b$, $V_{a}$, $V_{rb}$ and $V_{nb}$ and investigate how temperature profile and thermal conductivity $\lambda$ are affected. It was found that potentials $V_b$ and $V_a$ are strictly needed to avoid a collapse of the nanotube. Results corresponding to several setups are reported in Fig. 3 and Table 2. It is worth stressing that, for all simulations in a vacuum, nonbonded interactions $V_{nb}$ proved to have a negligible effect on both the slope of temperature profile and heat flux at steady state. On the contrary, the torsion potential $V_{rb}$ does have impact on the temperature profile while no significant effect on the heat flux was noticed: As a consequence, in the latter case, thermal conductivity shows significant dependence on $V_{rb}$. More specifically, the higher torsion rigidity the flatter the temperature profile. 
%%%%%%%%%%%

\section{Thermal boundary conductance of a carbon nanofin in water}
%%%%%%
\subsection{Steady state simulations}
%%%%%%
In this section, we investigate on the heat transfer between a carbon nanotube and a surrounding fluid (water). The latter represents a first step towards a detailed study of a batch of single carbon nanotubes (or small bundles) utilized as {\em carbon nanofins} to enhance the heat transfer of a surface when transversally attached to it. To this end, and limited by the power of our current computational facilities, we consider a ($5,5$) SWNT (with a length $L \le 14[nm]$) placed in a box filled with water (typical setup is shown in Fig. 4). SWNT end temperatures are set at a fixed temperature $T_{hot}=360[K]$, while the solvent is kept at $T_w=300[K]$. The carbon-water interaction is taken into account by means of a Lennard-Jones potential between the carbon and oxygen atoms with a parameterization ($\epsilon_{CO}$, $\sigma_{CO}$) reported in table 1. Moreover, nonbonded interactions between the water molecules consist of both a Lennard-Jones term between oxygen atoms (with $\epsilon_{OO}$, $\sigma_{OO}$ from table \ref{table.param}) and a Coulomb potential:
%%%%%%%%%%%%%%%%
\begin{equation}
V_c \left( r_{ij} \right) = \frac{1}{4 \pi \varepsilon_0} \frac{q_i q_j}{r_{ij}},
\end{equation}
%%%%%%%%%%%%%%%%
where $\varepsilon_0$ is the permittivity in a vacuum while $q_i$ and $q_j$ are the partial charges with $q_O=-0.82$ e and $q_H=0.41$ e (see also \cite{Koum01}).  

We notice that, the latter is a classical problem of heat transfer (pictorially shown in Fig. 5), where a single fin (heated at the ends) is immersed in fluid maintained  at a fixed temperature. This system can be conveniently treated using a continuous approach under the assumptions of homogeneous material, constant cross section $S$ and one-dimensionality (no temperature gradients within a given cross section) \cite{kreith.book}. In this case, both temperature field and heat flux only depend on the coordinate $x$, and the analytical solution of the energy conservation equation yields, at the steady state, the following relationship:
%%%%%%%
\begin{equation}\label{analytical.sol}
\tilde T \left( x \right) = M e ^ {- m x} + N e ^ {m x},
\end{equation}
%%%%%%%
where $\tilde T \left( x \right) = T \left( x \right) - T_w$ denotes the difference between the local temperature at an arbitrary position $x$ and the fixed fluid temperature $T_w$. Let $\alpha$ and $C$ be the thermal boundary conductance and the perimeter of the fin cross sections, respectively, $m$ is linked to geometry and material properties as follows:
\begin{equation}\label{m.param}
m = \sqrt{\frac{\alpha_{st} C}{\lambda S}},
\end{equation}
whereas the two parameters $M$ and $N$ are dictated by the boundary conditions, $T \left( 0 \right) = T\left( L \right) = T_{hot}$ (or equivalently, due to symmetry, zero flux condition: $dT/dx \left( L/2 \right) = 0$), namely:
%%%%%%%%%%%%%%%
\begin{equation}
M = \tilde T \left( 0 \right) \frac{e^{mL/2}}{e^{mL/2} + e^{-mL/2}}, \quad N = \tilde T \left( 0 \right) \frac{e^{-mL/2}}{e^{mL/2} + e^{-mL/2}}.
\end{equation}
%%%%%%%%%%%%%%%%%%%%%%
\begin{figure}
	\centering
		\includegraphics[width=0.65\textwidth]{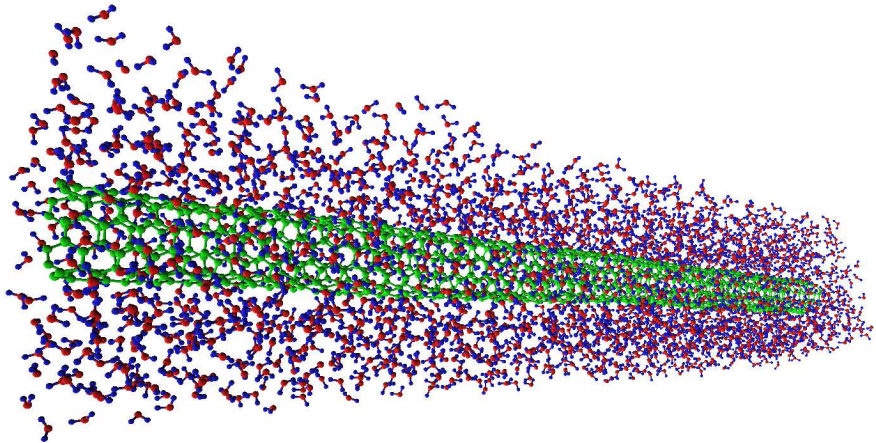}
	\caption{A ($5,5$) SWNT (green) is surrounded by water molecules (blue, red). Nos\'e-Hoover thermostats with temperature $T_{hot}=360[K]$ are coupled to the nanotube tips, while water is kept at a fixed temperature $T_w=300[K]$. After a sufficiently long time (here $15 [ns]$), a steady state condition is reached. MD simulation results (in terms of both temperature profile and heat flux) are consistent with a continuous one-dimensional model as described by eq. (\ref{temp.profile.exact}) and (\ref{heat.flux.exact}). Image obtained using VEGA ZZ \cite{VEGAZZ}.}\label{SWNTplusWATER}
\end{figure}
%%%%%%%%%%%%%%%%%%%%%%
%%%%%%%%%%%%%%%
Thus, the analytical solution (\ref{analytical.sol}) takes a more explicit form:
%%%%%%%%%%%%%%%
\begin{equation}\label{temp.profile.exact}
\tilde T\left( x\right) = \tilde T \left( 0 \right) \frac{\cosh{\left[ m \left( L/2-x\right)\right]}}{\cosh{\left( mL/2\right)}},
\end{equation}
whereas the heat flux at one end of the fin reads:
\begin{equation}\label{heat.flux.exact}
q_0 =   m \lambda S \tilde{T}\left( 0 \right) \tanh{ \left( mL/2 \right) }.
\end{equation}
%%%%%%%%%%%%%%%%%%%%%%
\begin{figure}
	\centering
		\includegraphics[width=0.45\textwidth]{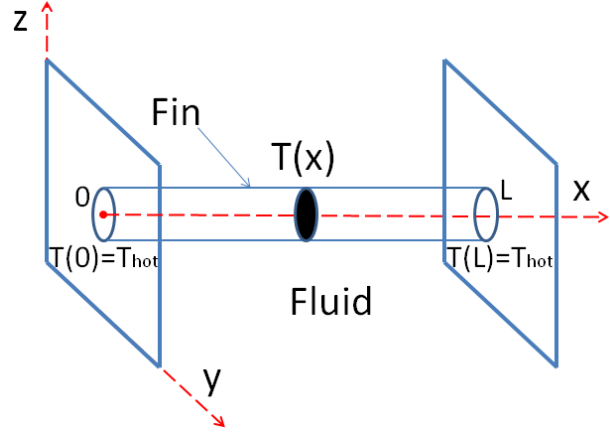}
	\caption{A single fin surrounded by a fluid can be studied by a one-dimensional continuous model, where all fields are assumed to vary only along the $x$-axis.}\label{schizzo-Fin}
\end{figure}
%%%%%%%%%%%%%%%%%%%%%%
In the setup illustrated in Fig. 4 and 5, periodic boundary conditions are applied in the $x$, $y$ and $z$ directions and all simulations are carried out with a fixed time step $dt = 1 [fs]$ upon energy minimization. First of all, the whole system is led to thermal equilibrium at $T=300$ by Nos\'e-Hoover thermostatting implemented for $0.8 [ns]$ with a relaxation time $\tau_T=0.1 [ns]$. Next, the simulation is continued for $15 [ns]$ where Nos\'e-Hoover temperature coupling is applied only at the tips of the nanofin (here, the outermost 16 carbon atom rings at each end) with $T_{hot}=360[K]$, and water with $T_{w}=300[K]$ until, at the steady state, the temperature profile in Fig. 6 is developed. Moreover, pressure is set to $1[bar]$ by Parrinello-Rahman pressostat during both thermal equilibration and subsequent non-equilibrium computation. We notice that the above molecular dynamics results are in a good agreement with the continuous  model for single fins if $mL/2=0.28$. Hence, this enables us to estimate the thermal boundary conductance $\alpha_{st}$ between SWNT and water with the help of eq. (\ref{m.param}):
\begin{equation}
\alpha_{st}=\frac{m^2 \lambda S}{C}.
\end{equation}
The thermal conductivity $\lambda$ has been independently computed by means of the technique illustrated in the sections above for the SWNT alone in a vacuum. Results for a nanofin with $L=14 [nm]$ are reported in Table 2.
%%%%%%%%%%%%%%%%%%%%%%%%%%%%%
%%%%%%%%%%%%%%%%%%%%%%
\begin{figure}
\centering
	\includegraphics[width=0.55\textwidth]{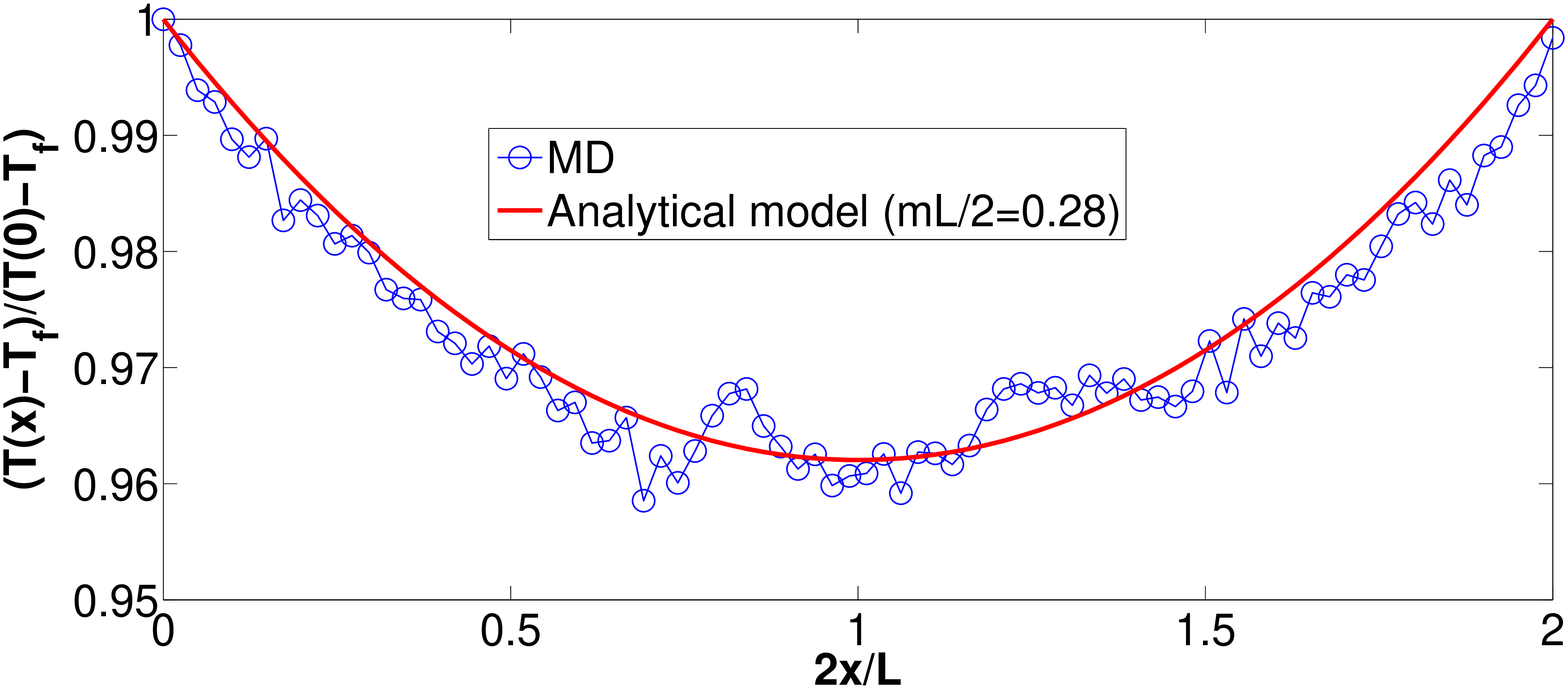}
	\caption{Steady state molecular dynamics (MD) simulations. Dimensionless temperature computed by MD (symbols) versus temperature profile predicted by continuous model (line), eq. (\ref{temp.profile.exact}). Best fitting is achieved by choosing $mL/2=0.28$. Case with computational box $2.5 \times 2.5 \times 14 [nm^3]$.}\label{temp.prof.nanofin}
\end{figure}
%%%%%%%%%%%%%%%%%%%%%%
We stress that heat flux computed by time averaging of the Nos\'e-Hoover parameter $\xi$ (see eq. (\ref{heat.flux.nh})) is also in excellent agreement with the value predicted by the continuous model through eq. (\ref{heat.flux.exact}). For instance, with the above choice $mL/2=0.28$, for ($5,5$) SWNT with $L=10[nm]$, $L_{NH}=2[nm]$ in a box $5 \times 5 \times 14 [nm^3]$ we have: $ - \left\langle \xi \right\rangle N_f k_b T = 3.11 \times 10^{-8} [W]$ while
\begin{equation}\label{flux.check}
 q_0 = m \lambda S \tilde{T}\left( 0 \right) \tanh{ \left( mL/2 \right) } = 3.14 \times 10^{-8}.   
\end{equation}
%%%%%%%%%%%%%%%%%%%%%%
We stress that $L_{NH}$ is the axial length of the outermost carbon atom rings coupled to a thermostat at each end of a nanotube.
%%%%%%%%%%%%%%%%%%%%%%
Finally, a useful parameter when studying fins is the {\em thermal efficiency} $\Omega$, expressing the ratio between the exchanged heat flux $q$ and the ideal heat flux $q_{id}$ corresponding to an isothermal fin with $T(x)=T(0)$, $\forall x \in [0,L]$ \cite{kreith.book}. In our case, we find highly efficient nanofins:
\begin{equation}
\Omega=\frac{q}{q_{id}}=\frac{m \lambda S \tilde{T}\left( 0 \right) \tanh{ \left( mL/2 \right) } }{ \alpha_{st} C \tilde{T}\left( 0 \right) L/2 } = \frac{\tanh{ \left( mL/2 \right) }}{mL/2} = 0.975.
\end{equation}

\subsection{Transient simulations}
The value of thermal boundary conductance between water and a single wall carbon nanotube has been assessed by transient simulations as well. Results by the latter methodology are denoted as $\alpha_{tr}$ in order to distinguish them from the same quantities ($\alpha_{st}$) in the above section. Here, the nanotube was initially heated to a predetermined temperature $T_{hot}$ while water was kept at $T_{w} < T_{hot}$ (using in both cases Nos\'e-Hoover thermostatting for $0.6 [ns]$). Next, an NVE molecular dynamics (ensemble where number of particle N, system volume V and energy E are conserved) were performed, where the entire system (SWNT plus water) was allowed to relax without any temperature and pressure coupling. Under the assumption of a uniform temperature field $T_{CNT}(t)$ within the nanotube at any time instant $t$ (Biot number $Bi < 0.1$), the above phenomenon can be modeled by an exponential decay of the temperature difference $(T_{CNT}-T_w)$ in time, where the time constant $\tau_d$ depends on the nanotube heat capacity $c_T$ and the thermal heat conductance $\alpha_{tr}$ at the nanotube-water interface as follows: 
\begin{equation}
\tau_d = \frac{c_T}{\alpha_{tr}}.
\end{equation}  
In our computations, following \cite{HUX03}, we considered the heat capacity per unit area of an atomic layer of graphite $c_T=5.6 \times 10^{-4} [J m^{-2}K^{-1}]$. The values of $\tau_d$ and $\alpha_{tr}$ have been evaluated in different setups, and results are reported in the table 2. It is worth stressing that values for thermal boundary conductance obtained in this study are consistent with both experimental and numerical results found by others for single wall carbon nanotubes within liquids \cite{HUX03,Shenogin04}. 
%%%%%%%%%%%%%%%%%%%%%%
\begin{figure}
	\centering
		\includegraphics[width=0.45\textwidth]{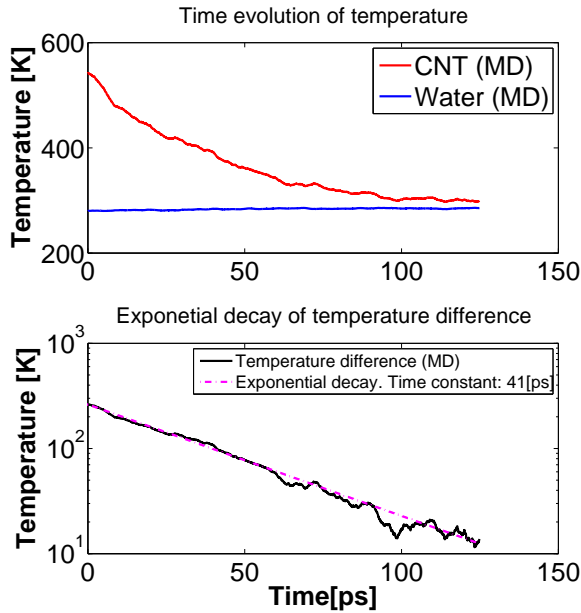}
	\caption{Transient simulations.	Temperature evolution as predicted by NVE molecular dynamics. Best fitting of exponential decay of the temperature difference $T_{CNT}-T_w$ is achieved by choosing: $\tau_d=41[ps]$.}\label{transient01}
\end{figure}
%%%%%%%%%%%%%%%%%%%%%%
%%%%%%%%%%%%%%%%%%%%%
 \begin{table}[htdp]\label{table.param}
\caption{Parameters for carbon-carbon, carbon-water and water-water interactions are chosen according to \cite{Guo91} and \cite{Koum01}.}
\begin{center}
   \begin{tabular}{|c|c|c|}
        \hline \multicolumn{2}{|c|}{Carbon-Carbon interactions}\\ \hline
        $k_{ij}^b$ & 47890 $kJ mol^{-1} nm^{-2}$ \\ \hline
        $k_{ijk}^\theta$ & 562.2 $kJ mol^{-1}$  \\ \hline
        $k_{ijkl}^\phi$ & 25.12 $kJ mol^{-1}$    \\ \hline
        $\epsilon_{CC}$ & 0.4396 $kJ mol^{-1}$    \\ \hline
        $\sigma_{CC}$ & 3.851 \AA    \\ \hline
        \hline \multicolumn{2}{|c|}{Carbon-Oxygen interactions}\\ \hline
        $\epsilon_{CO}$ & 0.3126 $kJ mol^{-1}$    \\ \hline
        $\sigma_{CO}$ & 3.19 \AA    \\ \hline
        \hline \multicolumn{2}{|c|}{Oxygen-Oxygen interactions}\\ \hline
        $\epsilon_{OO}$ & 0.6502 $kJ mol^{-1}$    \\ \hline
        $\sigma_{OO}$ & 3.166 \AA    \\ \hline
        \hline \multicolumn{2}{|c|}{Oxygen-hydrogen interactions}\\ \hline
        $q_{O}$ & -0.82 e    \\ \hline
        $q_{H}$ &  0.41 e    \\ \hline
      \end{tabular}
      \end{center}
%\label{table.param}
\end{table}
%%%%%%%%%%%%%%%%%%%%%%

\begin{table}[htdp]\label{table.conductivity}
\caption{Summary of the results of MD simulations in this work. Single wall nanotubes with chirality ($3,3$), ($5,5$) and ($15,0$) are considered, and several combination of interaction potentials are tested. In the first column, B, A, D and LJ stand for bond stretching, angular, dihedrals and Lennard-Jones potentials, respectively, while Bw denotes bond stretching with a smaller force constant $k_{ij}^b=42000 [kJ mol^{-1} K^{-1}]$ according to \cite{Brenner02}. Simulations are carried out both in a vacuum (vac) and within water (sol).}
\begin{center}
%%%%%%%%%%%%%%%%%%%%%%%%%%%%%%%%%
      \begin{tabular}{|c|c|c|c|c|c|c|c|c|}
        %\hline \multicolumn{5}{|c|}{ }\\ \hline
        \hline
        \hline
       Chirality, Case& Box & $L_{NH}$ & $L$ & $\lambda$ & $\alpha_{st}$ & $\alpha_{tr}$ & $\tau_d$ & $mL/2$\\
       & $[nm^3]$ & $[nm]$ & $[nm]$ & $\left[\frac{W}{m\,K}\right]$ & $\left[\frac{W}{m^2\,K}\right]$ & $\left[\frac{W}{m^2\,K}\right]$ & [$ps$] & \\
        \hline
        \hline
        ($5,5$), BAD-LJ (vac)  & $12 \times 12 \times 12$          & 1.5  & 5.5  & 67 & $-$ & $-$ & $-$ & $-$\\ \hline		               % $Wm^{-1}K^{-1}$  $[nm]$
        ($5,5$), BwAD-LJ (vac) & $12 \times 12 \times 12$          & 1.5  & 5.5  & 64 & $-$ & $-$ & $-$ & $-$\\ \hline
        ($5,5$), BAD    (vac)  & $12 \times 12 \times 12$          & 1.5  & 5.5  & 65 & $-$ & $-$ & $-$ & $-$\\ \hline
        ($5,5$), BA     (vac)  & $12 \times 12 \times 12$          & 1.5  & 5.5  & 49 & $-$ & $-$ & $-$ & $-$\\ \hline
        ($5,5$), BwA    (vac)  & $12 \times 12 \times 12$          & 1.5  & 5.5  & 48.9 & $-$ & $-$ & $-$ & $-$\\ \hline
        ($5,5$), BAD-LJ (vac)  & $20 \times 20 \times 20$    			 & $2$   &  $10$  & 96.9 & $-$ & $-$ & $-$ & $-$\\ \hline
        ($5,5$), BAD-LJ  (vac) &  $105 \times 105 \times 105$      & 25   & 25   & 216.1 & $-$ & $-$ & $-$   &   $-$\\ \hline
        ($5,5$), BAD-LJ (sol)  & $2.5 \times 2.5 \times 14$    & $2$   &  $10$  & $-$ & $5.18 \times 10^7 $ & $-$ & $-$ & $0.28$ \\ \hline
        ($5,5$), BAD-LJ (sol)  & $4 \times 4 \times 14$        & $2$   &  $10$ & $-$ & $5.18 \times 10^7 $ & $-$ & $-$ & $0.28$ \\ \hline
        ($5,5$), BAD-LJ (sol)  & $4 \times 4 \times 14$        & $0$   &  $14$ & $-$ & $-$ & $1.70 \times 10^7$ & $33$ & $-$ \\ \hline
        ($5,5$), BAD-LJ (sol)  & $5 \times 5 \times 5$         & 0      & 3.7  & $-$ &  $-$ & $1.37 \times 10^7$ & 41 & $-$  \\ \hline
        ($15,0$), BAD-LJ (sol)  & $5 \times 5 \times 5$        & 0      & 4.7  & $-$ &  $-$ & $1.60 \times 10^7$ & $35$   & $-$ \\ \hline
        ($15,0$), BAD-LJ (sol)  & $5 \times 5 \times 5$        & 0      & 3.8  & $-$ &  $-$ & $1.43 \times 10^7$ & $39$   & $-$ \\ \hline
        ($3,3$), BAD-LJ (sol)  &  $5 \times 5 \times 5$        & 0      & 3.7  & $-$ &  $-$ & $8.90 \times 10^6$ & $63$   & $-$ \\ \hline
      \end{tabular}
%%%%%%%%%%%%%%%%%%%%%%%%%%%%%%%%%%%
      \end{center}
%\label{table.conductivity}
\end{table}%
%%%%%%%%%%%
%%%%%%%%%%%
\section{Conclusions}
%  Text for this section \ldots
In this work, we first investigate the thermal conductivity of single wall carbon nanotubes by means of classical non-equilibrium molecular dynamics using both simplified one-dimensional and fully three-dimensional models. Next, based on the latter results, we have focused on the boundary conductance and thermal efficiency of single wall carbon nanotubes used as nanofins within water. More specifically, toward the end of computing the boundary conductance $\alpha$, two different approaches have been implemented. First, $\alpha=\alpha_{st}$ was estimated through a fitting procedure of results by steady state MD simulations and a simple one-dimensional continuous model. Second, cooling of SWNT (at $T_{CNT}$) within water (at $T_w$) was accomplished by NVE simulations. In the latter case, the time constant $\tau_d$ of the temperature difference $(T_{CNT}-T_{w})$ dynamics enables to compute $\alpha=\alpha_{tr}$. Numerical computations do predict pretty high thermal conductance at the interface (order of $10^7$ $[W m^{-2} K^{-1}]$), which indeed makes carbon nanotubes ideal candidates for constructing nanofins. We should stress that, consistently with our results $\alpha_{st} > \alpha_{tr}$, it is reasonable to expect that $\alpha_{st}$ represents the upper limit for the thermal boundary conductance, due to the fact that (in steady state simulations) water is forced by the thermostat to the lowest temperature at any time and any position in the computational box. Finally, it is useful to stress that, following the suggestion in \cite{Zhong06PRB}, all results of this work can be generalized to different fluids using standard nondimensionalization techniques, upon a substitution of the parameterization ($\epsilon_{CO}$, $\sigma_{CO}$) representing a different Lennard-Jones interaction between SWNT and fluid molecules.

\section{Authors contributions}
All one-dimensional atomistic simulations and numerical experiments for assessing thermal boundary conductances $\alpha$ were performed by E. Chiavazzo. Measurements of thermal boundary conductance through steady state ($\alpha_{st}$) and transient simulations ($\alpha_{st}$) were thought by P. Asinari and E. Chiavazzo respectively. Computations of thermal conductivity with different combination of interaction potentials, as reported in Fig. 3, were performed by P. Asinari. Authors contibuted equally in writing the present manuscript.

\section{Acknowledgments}
The research leading to these results has received funding from
the European Community Seventh Framework Program (FP7 2007-2013) under grant agreement N. 227407-Thermonano.
The Authors wish to state their appreciation to Dr. Marco Giardino for helping us all times we had troubles with our computational facilities. We thank Dr. Andrea Minoia and Dr. Thomas Moore for the fruitful discussions on the usage of GROMACS in simulating carbon nanotubes. We acknowledge interesting discussions with Dr. Jean-Antoine Gruss (CEA DTS/LETH, France) about CNT nanofluids.

\section{Methods}
The carbon nanotubes geometries simulated in this paper were generated using the program Tubegen \cite{tubegen}, while water molecules were introduced using the {\em SPC/E} model implemented by the {\em genbox} package available in GROMACS \cite{gromacsweb}.  
Numerical results in this work are based on non-equilibrium molecular dynamics where the all-atom forcefields OPLS-AA is adopted for modeling atom interactions. Visualization of simulation trajectories is accomplished using VEGA ZZ \cite{VEGAZZ}.

%\bibliographystyle{plain}
%\bibliography{bmc_article}

\end{document}